\input harvmac
\Title{\vbox{\hbox{HUTP--96/A057}
\hbox{IASSNS-HEP-96/125}\hbox{hep-th/9612052}}}
{\vbox{\centerline{F-theory, Geometric Engineering and N=1 Dualities}}}
\vskip .05in
\centerline{\sl M. Bershadsky$^{\natural}$,
A.  Johansen$^{\natural}$,
T. Pantev$^{\forall}$,
V. Sadov$^{\sharp}$ and
C. Vafa$^{\natural}$}
\vskip .2in
\centerline{\it $^{\natural}$ Lyman Laboratory of Physics, Harvard
University}
\centerline{\it Cambridge, MA 02138, USA}
\vskip .2in
\centerline{\it $^{\forall}$ Department of Mathematics, Masachusetts
Institute of Technology}
\centerline{\it Cambridge, MA 02138, USA}
\vskip .2in
\centerline{\it $^{\sharp}$ Institute for
Advanced Study}
\centerline{\it Princeton, NJ 08840, USA}

\vskip .2in
We consider geometric engineering of $N=1$ supersymmetric
QFTs with matter in terms of a local model for compactification
of F-theory on Calabi-Yau fourfold.  By brining 3-branes
near 7-branes we engineer $N=1$  supersymmetric $SU(N_c)$
gauge theory with $N_f$ flavors in the fundamental.  We identify
the Higgs branch of this system with the instanton moduli
space on the compact four dimensional
space of the 7-brane worldvolume.  Moreover we show that
the Euclidean 3-branes wrapped around the compact part
of the 7-brane worldvolume can generate
superpotential for $N_f=N_c-1$ as well as lead to quantum corrections
to the moduli space for $N_f=N_c$.  Finally we argue that Seiberg's
duality for $N=1$ supersymmetric QCD may be mapped
to T-duality exchanging 7-branes with 3-branes.

\vskip .2in
\noindent

\Date{ December 1996}

%\draft

\newsec{Geometric Engineering of SUSY QCD}

Many non-trivial aspects of field theory dualities in four dimensional
supersymmetric systems have found a natural interpretation in the context
of string theory. This is the case in particular for N=4
 and N=2
supersymmetric theories where, by a suitable engineering of
a local model for the compactification manifold
 the non-trivial dualities
of field theories can be reduced to an application of T-duality
(see e.g. \ref\geng{S. Katz, A. Klemm and  C. Vafa,
``Geometric Engineering of Quantum Field Theories'', hep-th/9609239.}\
for the $N=2$ case).
The case of N=1 theories has been
the most difficult case to study, though there has been some partial
progress. In this paper we find a natural
geometric engineering of $N=1$ quantum field theories
with matter in four dimensions
which again maps $N=1$ dualities  \ref\seib{K. Intriligator and  N. Seiberg,
``Lectures on supersymmetric gauge theories and electric-magnetic duality'',
Nucl. Phys. Proc. Suppl.  {\bf 45 BC} (1996) 1.}\
to T-dualities of string theory.

The basic approach is via F-theory, where we consider compactification
on elliptic Calabi-Yau 4-folds from 12 dimensions down to 4.
This leads to $N=1$ quantum field theories in four dimensions.
%\ref\mukhi{R. Gopakumar  and  S. Mukhi, Nucl. Phys. {\bf B 479} (1996) 260.}.
  In particular
a situation with pure N=1 Yang-Mills in $d=4$ was geometrically
engineered in \ref\kvg{S. Katz and  C. Vafa,
``Geometric Engineering of N=1 Quantum Field Theories'',
hep-th/9611090.}.
This was done by constructing
a 7-brane over which the elliptic fibration has ADE singularity.
The question we mainly
concentrate in this paper is a way to incorporate matter
into this system.  One way of doing this is in principle
to have extra singularities over the space
as in the  $N=2$ case \ref\kucha{M. Bershadsky,  K. Intriligator,  S. Kachru,
D. R. Morrison,  V. Sadov and C. Vafa,
``Geometric Singularities and Enhanced Gauge Symmetries'',
hep-th/9605200.}\
\ref\matterKV{S. Katz and  C. Vafa, ``Matter From Geometry'',
 hep-th/9606086.}.
There seems
to be certain obstacles in this way of getting matter in the N=1 theories,
because almost inescapably when one attempts to put extra
singularities they are of the form which does not seem to correspond
to conventional physics \ref\kvu{S. Katz, private communication.}\
(suggesting exotic
physics as in \ref\matterBJ{M. Bershadsky and  A. Johansen,
``Colliding Singularities in F-theory and Phase Transitions'',
hep-th/9610111.}).  Here we attempt
a different way of incorporating matter into the system.

As was noted in \ref\svw{S. Sethi ,  C. Vafa and  E. Witten,
``Constraints on Low-Dimensional String Compactifications'',
hep-th/9606122.}\ the F-theory on Calabi-Yau 4-fold
induces a certain number of 3-brane charge
which needs to be cancelled in order to get a consistent
theory.  One way of doing this is by adding a certain
number of 3-branes which fill the spacetime.  The number of 3-branes
needed for this is $\chi/24$ where $ \chi$ is the Euler characteristic
of Calabi-Yau fourfold (see also \ref\bb{  K. Becker and M. Becker,
``M-Theory on Eight-Manifolds'', Nucl. Phys. {\bf B 477} (1996) 155.}).
Now suppose we have a 7-brane which has pure ADE singularity,
without extra singularities. Note that the 7-brane worldvolume
is $R^4\times S$ where $S$ is a complex 2 dimensional surface
in the base of the elliptic 4-fold.
In general, in addition to $N=1$ Yang-Mills, we get
$h^{2,0}(S)+h^{1,0}(S)$ adjoint chiral
fields \kvg .  We will first consider the case where
we do not have any adjoint fields and comment on generalization
for the case with adjoints later in the paper.
This means that we first consider the case
\eqn\cond{h^{1,0}(S)=h^{2,0}(S)=0}
This assumption, together with the simplifying assumption of
rigidity of $S$ implies that $S$ is a rational surface.
 This local model  leads to pure ADE Yang-Mills in $d=4$ with no matter.
Let us concentrate on the case of $A_{N_c-1}$ singularity
which gives rise to $SU(N_c)$ Yang-Mills. In order to
obtain pure field theory results (i.e.
turning off gravitational/stringy effects) we need to consider
the limit where the volume of $S$ is very large.
 Note
that the position of the 3-branes
which fill the spacetime can vary over the three
dimensional base.  Suppose
we bring one of the 3-branes near the 2-dimensional surface $S$.
In this way we can analyze the result by a local analysis, which
leads to having a hypermultiplet in the fundamental of $SU(N_c)$
(which is also charged under the $U(1)$ charge on the 3-brane)
\ref\besv{M. Bershadsky ,  V. Sadov and  C. Vafa, ``D-strings on D-Manifolds'',
Nucl. Phys.  {\bf B 463} (1996) 398.}.
In terms of $N=1$ matter, this correponds to quark fields $Q$ and $\tilde Q$
in the representations ${\bf N_c}$ and ${\overline{\bf N_c}}$.  Moreover out
of the 3 chiral fields which correspond to moving the 3-brane
around, one of them, which takes the 3-brane off of $S$, can
give mass to the quark field.  If we bring $N_f$ of the 3-branes
near $S$ we obtain an $N=1$ Yang-Mills theory with
gauge group $SU(N_c)$ and with $N_f$ flavors $Q,{\tilde Q}$.

This system has two branches  \ref\seib{N. Seiberg,
``Electric-Magnetic Duality in Supersymmetric Non-Abelian Gauge Theories''
Nucl. Phys. {\bf B 435} (1995) 129.}.
  On the
one hand we can give mass to the quarks, which as noted above
corresponds to taking the 3-branes off the 7-brane surface $S$.
There is also a Higgs branch. Note that if $N_f< N_c-1$ maximal
Higgsing will leave us with $SU(N_c-N_f)$ unbroken gauge symmetry
with no matter.  On the other hand when
 $N_f\geq N_c-1$ we can
higgs the group completely.
The classical moduli space of the Higgs branch is obtained by
considering the gauge invariant observable made out of
$Q,\tilde Q$.  The mathematical way of saying this
is that the moduli space of Higgs branch is simply
\eqn\modu{{\cal M}={{\bf C}^{2N_fN_c}\over {SL(N_c)}}}
where the complex space ${\bf C}^{2N_fN_c}$ corresponds
to expectation values for $Q,\tilde Q$ and the quotient
represents the gauge invariant observables (see e.g. \ref\tay{
M. A. Luty and  W. Taylor IV, ``Varieties of vacua in classical
supersymmetric gauge theories'', Phys. Rev. {\bf D 53} (1996) 3399.}).
For $N_f>N_c$ the complex structure of this moduli
space does not receive quantum corrections
whereas for $N_f=N_c$ it receives quantum corrections \ref\seiberg{N. Seiberg,
Phys. Rev.
{\bf D 49} (1994) 6857.}.
Note that the complex dimension of the moduli space when complete
Higgsing is possible is given by
\eqn\dim{{\rm dim}{\cal M}^{Higgs}=2N_fN_c-(N_c^2-1)}

The question is how this branch of the moduli space is realized
in the context of F-theory under discussion.  The answer is very
simple, once we note that 3-branes within 7-branes can be
equivalently viewed as corresponding to zero size instantons
of the corresponding 7-brane
\ref\small{E. Witten, Nucl. Phys.
{\bf B460} (1996) 541.}\
\ref\doug{M. Douglas, {\it Gauge Fields and
D-branes},
hep-th/9604198.}\ \ref\vin{C. Vafa, ``Instantons on D-branes'',
Nucl. Phys. {\bf B463} (1996) 435.}.
If we consider finite size instantons, we break the $SU(N_c)$ gauge symmetry
 and will thus Higgs the system\foot{
Aspects of this transition has been considered recently
in connection with $N=1$ F-theory/heterotic dualities
in \ref\BJPS{M. Bershadsky, A. Johansen, T. Pantev and V. Sadov,
``On Four-Dimensional Compactifications of   F theory'',
to appear.}.}.  Thus {\it we identify}
the Higgs branch of this system with the moduli of
$SU(N_c)$ instantons on $S$ with instanton number $N_f$.
As a first check let us see what the dimension of this moduli space
is.  For any gauge group $G$ on a four dimensional space $S$
with instanton number $k$ the complex dimension of moduli space is
(assuming instantons exist)
\eqn\dimf{{\rm dim}{\cal M}^{inst}=2kc_2(G)-dim(G){[\chi +\sigma]\over 4}
=2kc_2(G)-dim(G)(h^{2,0}-h^{1,0}+h^{0,0})}
where $c_2(G)$ denotes the dual coxeter number of the group $G$
and $\chi$ and $\sigma$ denote the Euler characteristic
and the signature of $S$ respectively whose sum is related to the hodge
numbers of $S$ as indicated above.
 For the case at hand $k=N_f$
and $c_2(SU(N_C))=N_c$, and we are taking $h^{1,0}=h^{2,0}=0$.  Thus we find
that the
dimension of instanton moduli space is
$${\rm dim}{\cal M}^{inst.}=2N_fN_c-(N_c^2-1)$$
in agreement with \dim .  This formula is valid for $N_f\geq N_c-1$
(there is some subtlety for $N_f=N_c-1$ noted below).
Not only the dimensions match, but one can argue that in fact
the moduli space is identical to the expectation based on the
gauge theory realization \modu.

This comes from a well known mathematical
construction \ref\wei{Wei-Ping Li and Zhenbo Qin, {\it
Stable vector bundles on algebraic
   surfaces}, Trans. Amer. Math. Soc. {\bf 345} (1994), no. 2, 833.}
for instantons on rational surfaces which we now review.
It can be shown that $SU(N_c)$ instantons on $S$ exist
if and only if the instanton number $N_f$ satisfies
$N_f\geq N_c$. (For the case of $N_f=N_c-1$
we can only define an $SU(N_c-1)$ instanton.  This special
case will be treated later.)   Note that these facts are in accord with the
cases
of complete Higgsing expected from the field theory
analysis.  Now let us consider the detailed construction.
We will always be interested in a piece of the moduli of instantons on
a rational surface which is a neighborhood of an instanton of zero size.
Geometrically this amounts to looking at the restriction of
instantons on the complement of finitely many curves on the surface. Since
all rational surfaces become isomorphic after we throw out some curves we
have the freedom to choose a particular model of our surface.
For simplicity we concentrate on the case where
$S$ is a Hirzebruch surface.  For example we can take
$S={\bf P}^1\times {\bf P}^1$.   More generally we consider $S$ to be a
${\bf P}^1$ bunlde over ${\bf P}^1$ given by
$\pi: S \rightarrow {\bf P}^1$.
Let us denote the moduli space of
vector bundles of rank $N_c$ with vanishing first Chern class
 and second Chern class equal to $N_f$
by ${\cal M}_S(N_c,0,N_f)$.  Below we will describe an open dense subset in
${\cal M}_S (N_c,0,N_f)$.  First we fix a special vector bundle $W$ over $S$
of rank $N_c$ and zero instanton number, $c_2(W)=0$.
The instanton number can be changed by
a procedure known as the {\it Hecke transform} which
we will describe below.  We consider
a divisor $D=\sum_1 ^{N_f} y_i$ consisting of $N_f$ distinct
${\bf P}^1$ fibers
of $\pi$.
A generic vector bundle is realized as a Hecke transform of $W$ along divisor
$D$.    Each Hecke transform will increase the instanton number by 1
and can roughly be associated with giving a zero size instanton
a finite size.
Since these Hecke transforms occur along the $N_f$ fibers the resulting
 bundle will have second Chern class
equal to $N_f$.   All moduli of the vector bundles are encoded in the
deformations of the Hecke transform.

We briefly review the definition of a Hecke transform. The
Hecke transform of a vector bundle $W\rightarrow
S$ along a divisor $D \subset S$ depends on the additional choice
of a vector bundle $F$ on $D$ and a surjective map of vector bundles $\xi :
W_{|D} \rightarrow F$. This additional data is called
a center of the Hecke transform. Given $W$, $F$ and $\xi$ define
the Hecke transformed bundle  $\widetilde{W}$ of $W$ by the exact sequence
\eqn\heck{
0 \longrightarrow \widetilde{W} \longrightarrow W  \longrightarrow F
\longrightarrow 0,
}
where the map $W \rightarrow F$ is given by $\xi$. It turns out that
$\widetilde{W}$ is a locally free sheaf of the same rank as
$W$. Moreover $\widetilde{W}$ coincides with $W$ everywhere
except along the divisor $D$. Sometimes the Hecke transform is called an
elementary
modification of $W$. For more details see \ref\mar{M. Maruyama,
Elementary transformations in the theory of
algebraic vector bundles. Algebraic geometry (La Rabida, 1981), 241--266,
Lecture Notes in Math., 1961.}.

Now let us come back to the situation at hand.
We first wish to define the bundle $W$.  Any rank $N_c$ bundle
over ${\bf P}^1$ can be written as a sum of line bundles.
Consider the integers $a$ and $b$ defined by $N_{f} = a N_{c} + b$ where
$0 \leq b < N_c$. We define the bundle $W$ over $S$ to be the pull-back
$$
W = \pi^{*}\left( {\cal O}_{{\bf P}^{1}}(a)^{\oplus (N_c-b)}\oplus
{\cal O}_{{\bf P}^{1}}(a+1)^{\oplus b} \right).
$$
The divisor $D$ is a collection of $N_f$ copies of ${\bf P}^{1}$ and
we choose $F$ to be the line bundle ${\cal O}(1)$ on each of
these ${\bf P}^{1}$'s.  Note that ${\cal O}(1)$ has two sections.
The restriction $W_{|D}$ is the trivial
rank $N_c$ vector bundle on $D$. In order to specify a Hecke transform of $W$
we need also a surjective map $\xi : W_{|D}\rightarrow  F$. The moduli of all
such
$\xi$'s are
\eqn\hom{
Hom(W_{|D}, F) = \oplus_{i = 1}^{N_f} Hom({\cal O}^{\oplus N_c}_{y_{i}},
{\cal O}_{y_{i}}(1)) = \oplus_{i = 1}^{N_f} H^{0}(y_{i},
{\cal O}_{y_{i}}(1)^{\oplus N_c}).
}
Since each fiber $y_{i}$ is a ${\bf P}^{1}$ and the global sections
$H^{0}({\bf P}^{1}, {\cal O}(1))$ are just the two homogeneous
coordinates, the right hand side of \hom\ is simply a complex vector space of
dimension $2N_c N_f$ and we identify $\xi=(Q,{\tilde Q} )$ squark
fields, where each squark field goes into defining
a Hecke transform on each component $y_i$ of the divisor $D$.
 Two different maps $\xi$ and $\xi'$ lead to the same
vector bundle whenever they differ by automorphisms of the bundle $W$ and
the sheaf $F$. The inverse is also true \wei . Since the $k$-tuples of points
on ${\bf P}^{1}$ are parametrized by a $N_{f}$-dimensional vector space
that can be thought of as a quotient of ${\bf C}^{N_{f}+1}- \{ 0 \}$ by ${\bf
C}^{*}$ we can describe the full instanton moduli
${\cal M}_S(N_c,0,N_f)$ as a quotient of an open set in
${\bf C}^{2N_{c}N_{f}}\oplus {\bf C}^{N_{f}}$ by the group $Aut(W)/{\bf C}^{*}
\times Aut(F) \times {\bf C}^{*}$.  This description of moduli
space is dual to the one given in \wei .
The automorphism group of $F$ is ${\bf C}^{*k}$. Since we are interested in the
limit of the moduli space when the size of $S$ is growing to infinity we need
the corresponding limit of this quotient. But in this limit the base ${\bf
P}^{1}$ goes to
infinite size and hence the construction will remember only the restriction of
$W$ on the complement of a fiber of $\pi$ and hence $Aut(W)/{\bf C}^{*} $
reduces to  $SL(N_{c},{\bf C})$. In particular, the actions of $Aut(W)/{\bf
C}^{*}$ and ${\bf C}
^{*k}$ decouple and in the limit the moduli looks like ${\bf C}^{2N_{c}N_{f}}/
SL(N_{c})$.

\newsec{Quantum Moduli Space}

So far we have only discussed how the classical
aspects of the gauge system can be engineered in the context
of F-theory compactification on 4-folds.
As was argued in \seiberg\ the classical moduli space
receives a quantum correction only for $N_f=N_c$, where
if we consider the $N_f\times N_f$ meson field $M=Q\tilde Q$
and let $B=det Q$ and $\tilde B=det {\tilde Q}$ denote
the two baryon fields, then the quantum moduli space is
$${\rm det}M-B\tilde B=\Lambda^{2N_c}$$
where $\Lambda^{2N_c}=\mu^{2 N_c}{\rm exp}({-8\pi^2\over g^2})$
comes from a point-like instanton configuration.  Moreover for
the case $N_f=N_c-1$ one expects a superpotential due
to point-like instantons \ref\ads{I. Affleck, M. Dine and N. Seiberg,
Nucl. Phys. {\bf B 241} (1984) 493; Nucl. Phys. {\bf B 256} (1985) 557.}\
$$W={{\tilde \Lambda}^{2N_c+1}\over det M}$$
We now wish to look for such corrections in the present setup.
The basic instanton to consider
can be geometrically understood in the present context by noting
that a point like instanton in uncompactified spacetime
for the 7-brane, corresponds to a Euclidean 3-brane wrapping
around $S$, whose action goes as $\sim {\rm exp}(-V)$
where $V$ is the volume of $S$. Moreover the volume of $S$
is indeed ${1\over g^2}$ (from the reduction of the gauge
theory from 8 dimensions down to 4 on $S$). This is thus
in agreement with the form of the correction. To argue
that there is such a correction, we will have to do
the zero mode analysis for this euclidean three brane
similar to what was done for Euclidean membranes \ref\sbb{
K. Becker,  M. Becker and A. Strominger,
``Fivebranes, Membranes and Non-Perturbative String Theory'', Nucl. Phys. {\bf
B 456} (1995) 130.}\
 and euclidean fivebranes  \ref\witf{E. Witten,
``Non-Perturbative Superpotentials In String Theory'', Nucl.Phys.
{\bf B 474} (1996) 343.}.  The analysis
relevant for us is a simple extension
of \witf .  The main new novelty  here
is that when a Euclidean 3-brane wraps
$S$ this
will lead to a hypermultiplet in the fundamental of $SU(N_c)$
corresponding to the open string stretched between
the Euclidean 3-brane and the $(N_c)$ 7-branes.  This mode
propagates on the euclidean worldvolume $S$
of the 3-brane.  The reader should be careful to distinguish
between the $N_f$ fundamentals of $SU(N_c)$ which propagate
on uncompactified spacetime and the one fundamental of $SU(N_c)$
which lives on $S$.
  We consider
a particular point on the Higgs moduli, which corresponds to some fixed
$SU(N_c)$ instanton on $S$, with instanton number $N_f$
(i.e. giving vevs to the spacetime squarks).  The hypermultiplet
living on the 3-brane worldvolume propagates on this fixed background.
The Euclidean 3-brane worldvolume theory is
twisted \ref\bsv{  M. Bershadsky ,  V. Sadov and  C. Vafa, ``D-Branes and
Topological Field Theories'',
Nucl. Phys.  {\bf B 463} (1996) 420.}\ and in this case the twisting
is the one considered in \ref\joh{A. Johansen, Int. J. Mod. Phys. {\bf A 10}
(1995) 4325.}.  This twisting
leads to fermions in the hypermultiplet
being represented by anti-holomorphic forms $\Omega^{0,0}\oplus
\Omega^{0,1}\oplus \Omega^{0,2}$ with coefficients in the fundamental
representation of $SU(N_c)$.  Just as in \witf\ we can define
a conserved charge $W$ which basically corresponds to the twisting
of the theory along $S$.  If we consider the charge carried
by the fermions it is $\pm {1\over 2}$ correlated with the
degree of the form (i.e. the chirality of the spinor before
twisting).
 The net violation of this charge due to fermion zero modes
(which is doubled because of the taking into account both
components of a hypermultiplet), is given
by the index of $\overline \partial_A$ where $A$
denotes the $SU(N_c)$ gauge connection on $S$.  The index
for this complex is found by a simple application
of Atiyah-Singer index theorem to be
given by
\eqn\vio{\Delta W=ind({\overline \partial}_A)=n_0-n_1+n_2=
N_c-N_f}
where $n_i$ denotes the number of holomorphic section of degree
$i$-antiholomorphic forms with coefficient in the fundamental representation
of $SU(N_c)$.  As argued in \witf\ in order for the instanton
to contribute to the superpotential there must be a violation
of $\Delta W =1$ (corresponding to the charge
of $d^2\theta$ and thus a contribution to the superpotential).
In the same way one would expect that if we want a correction
to the complex moduli of the vacua we should have zero net violation
of $W$, i.e., $\Delta W=0$ \ref\witpc{E. Witten, private communication.}.
This would in particular lead to a non-vanishing contribution
from the instanton without involving the integration over $d^2\theta$.
Since this is an instantonic contribution it should only affect holomorphic
quantities and this should thus lead to quantum correction to the
moduli space.   From the formula \vio\ we see that the case
corresponding to quantum correction to the moduli space is
$N_f=N_c$ and the case corresponding to generation
of superpotential is $N_f=N_c-1$.  As noted in the
quantum field theory setup, the case $N_f=N_c$ is indeed
expected to lead to quantum correction to the complex
structure of moduli space, confirming the above analysis.
Note also that this contribution is of the form ${\rm exp}(-V)
={\rm exp}(-1/g^2)$ given the relation between the volume of $S$ and
the gauge coupling constant.
Also the case with $N_f=N_c-1$ is the case where one expects
point-like instanton contributions to the superpotential.
However we should analyze this case more carefully, because
our analysis for instantons applied only to the cases where
$N_f\geq N_c$.

Let us now consider the $N_f=N_c-1$ case
in detail.  In this case as mentioned before
we cannot have $SU(N_c)$ instantons
on $S$, but we can have $SU(N_c-1)$ instantons on it.
This will in effect describe the Higgs branch in this case.
The reason for this is that giving vev to squarks
in the $SU(N_c-1)$ subgroup of the color group
effectively Higgses the group completely.
Let us see if this is also in agreement
with the dimension of moduli of $SU(N_c-1)$ instantons with
instanton number $SU(N_c-1)$.  The dimension of instanton moduli
space in this case is
$${\rm dim}{\cal M}^{inst.}=(N_c-1)^2+1$$
However taking into account that with an $SU(N_c-1)$
instanton, we still have an $U(1)$ symmetry
of the original $SU(N_c)$ theory left, this acts by a $C^*$ action
on the above moduli space leaving us with the reduced moduli space
of dimension $(N_c-1)^2$.  This is easily verified to be
the dimension of the Higgs branch expected in this case.
Now let us return to the contribution of the Euclidean 3-brane
instanton in this case.  The 1 fundamental hypermultiplet
propagating on $S$ will be decomposed into a fundamental
of $SU(N_c-1)$ plus a singlet.  The net violation
of $W$ will come only from the singlet component, in which
case
$$\Delta W=n_0-n_1+n_2=h^{0,0}-h^{0,1}+h^{0,2}=1$$
We will thus get a correction to the superpotential as expected.
To find what the superpotential contribution of the Euclidean
3-brane is, in addition to the instanton action which leads
to the ${\rm exp}(-1/g^2)$ we will get contribution from the
determinant of
non-zero modes of the fields in the instanton background \witf .
In this case we have to also take into account the bosonic
components of the hypermultiplet, which belong to degree
$0$ and $2$ anti-holomorphic forms coupled to the fundamental
representation of $SU(N_c-1)$.  Putting the contribution
of non-zero modes for bosons and fermions one finds
that the prefactor is precisely the Ray-Singer torsion\foot{
The Ray-Singer torsion in twisted N=1 SUSY theories has been first
considered in \joh .}
of the $\overline \partial_A$, which is given by
$$RS=\prod_p {\rm det} \Delta_p^{-(-1)^pp}$$
where $\Delta_p$ denotes the laplacian acting
on anti-holomorphic p-forms with coefficient in the fundamental
of $SU(N_c-1)$.  The Ray-Singer torsion $RS$ will depend
on the complex moduli of the instanton.  Given the fact
that the instanton moduli space is $SU(N_f)$ symmetric,
this can only be a function of $\psi={\rm det} M$.  Now consider
the limit where we turn off one of the instantons.  This
means that we
consider a point where $M$ has one lower rank that the maximal rank of
$N_f=N_c-1$.  In other words this corresponds to the point where $\psi =0$.
In this case we find that there is an extra bosonic zero mode
contribution to the Ray-Singer torsion.  This follows from the
fact that the index of $\overline \partial_A$ goes up by one
when we shrink one instanton and the RS torsion has the same
order of pole/zero as this index. This implies that as $\psi \rightarrow 0$
we get an extra pole in the superpotential.  This implies
that the RS torsion behaves as
$$RS={1\over \psi}+...$$
where the $...$ are non-singular.  However since the field theory
limit is obtained in the limit where the size of $S$ goes to infinity,
or equivalently the $M\rightarrow 0$ the subleading
terms are suppressed by powers of Planck constant.  Thus we would
expect a superpotential of the form
$$W={1\over {\rm det}M} {\rm exp}{-1\over g^2}$$
in perfect accord with field theoretic expectations\foot{
Morally the Ray-Singer torsion is a holomorphic function of $\psi$,
but as found in cases studied in \ref\bcov{M. Bershadsky, S. Cecotti,
H. Ooguri and C. Vafa, Comm. Math. Phys {\bf 165} (1994) 311.}\
one expects holomorphic anomalies for this quantity.  However
the poles can be shown to always be holomorphic functions.}.  In fact
we can derive this more directly using the D-brane techniques.
Consider the limit where we turn off all the instantons.  In this
case we have $N_f$ 3-branes which fill the space.  In the
presence of the Euclidean
threebrane we get the additional bosonic modes $\alpha^i, {\tilde \alpha}_i$
coming from open string of zero length stretched between
the Euclidean 3-brane and the $N_f$ physical 3-branes.  Moreover
one finds a four point interaction of fields realized by open strings
between the physical 3-branes, the 7-brane and the Euclidean
3-brane, of the form
$$S=Q_i\tilde Q^j \alpha^i \tilde \alpha_j$$
To leading order this leads to the contribution to the instantons
of the form
$$\int d \alpha^i d\alpha_j {\rm exp}(Q_i\tilde Q^j \alpha^i \tilde \alpha_j)=
{1\over {\rm det}Q_i\tilde Q^j}.$$
which is as expected.

\newsec{Seiberg's $N=1$ Duality}
One of the striking aspects of the system under consideration
is that there is a dual description of it in terms of a different
gauge group $SU(N_f-N_c)$ discovered by Seiberg \seib.
How can one understand this dual
description in the present context?  We will argue
here that this should follow from applying T-duality to the
present system, which maps 7-branes to 3-branes and 3-branes
to 7-branes, by doing the analog of $R\rightarrow 1/R$ along
the four dimensional space represented by $S$.

Let us first consider the simpler case where $S=K3$.
In this case, instead of getting an $N=1$ system
we obtain an $N=2$ system with $N_f$ hypermultiplets
in the fundamental.  The situation can be described
in the perturbative type IIB setup, where we have $N_c$
parallel Dirichlet 7-branes on $R^4\times K3$.
We also have $N_f$ Dirichlet 3-branes filling $R^4$
and corresponding to $N_f$ points on $K3$.
This does not mean that the net 3-brane charge
is just $N_f$.  This is because the curvature
of $K3$ induces $-1$ unit of 3-brane charge for each 7-brane
\bsv
\ref\hgm{ M. Green, J. Harvey, G. Moore,
``I-Brane Inflow and Anomalous Couplings on D-Branes'',
hep-th/9605033.}.  Since we have
($N_c$) 7-branes we should get a $-N_c$ contribution
to the 3-brane charge.  This shift in lower D-brane
charge was crucial in checking the consequences of
string-string duality \bsv \vin .
Taking into account the $N_f$ 3-branes which
we have introduced we find that we have a net $N_f-N_c$
3-brane charge.  Now we apply a T-duality on $K3$ which
is a total inversion of the volume of $K3$.  This in particular
map the 7-brane charges to 3-brane charges and vice-versa.
One way to realize this is to consider $K3$ as an orbifold
$T^4  / {\bf Z}_2$ and apply the usual
$R\rightarrow 1/R$ duality
on each of the 4 circles of $T^4$.  We now end up
with $N_f-N_c$ 7-branes which fill the dual $K3$ times
the uncompactified spacetime.  We now
must have the net 3-brane charge of $N_c$.  Since we
now have $N_f-N_c$ 7-branes on the dual $K3$ which
induce $-N_f+N_c$ the brane charge, we must thus
explicitly have $N_f$ threebranes in addition.  Thus
under the T-duality we find the dual gauge group to be
$SU(N_f-N_c)$ again with $N_f$ fundamental hypermultiplets.
This thus shows that the higgs branch of $N=2$ theories
with $ SU(N_c)$ gauge group with $N_f$ fundamental
hypermultiplets should be the same as that of $SU(N_f-N_c)$
again with $N_f$ fundamental hypermultiplets!  This
has already been noted in \ref\ant{I. Antoniadis and
B. Pioline, ``Higgs branch, hyperk\"ahler quotient and duality in SUSY N=2
Yang-Mills theories'', hep-th/9607058.}.
Note that in the above if we had used $S=T^4$ instead of $K3$
a similar consideration would have applied, except that
we would have additional adjoint hypermultiplets, and that
the duality would simply exchange $N_f\leftrightarrow N_c$
without a shift in $N_f$, because there is no lower
D-brane charge induced on flat $T^4$.  In this
case the duality is the well known duality of the exchange
of instanton number with the rank of the group, and is
known as the Fourier-Mukai transform
\ref\muk{S. Mukai, Nagoya Math. J. {\bf 81} (1981) 153.}.

Now we come to the $N=1$ case.  We will now argue
that under the T-duality applied to $S$ the rank of the dual
gauge group and the number of hypermultiplets is exactly
as in the $N=2$ case discussed above where we took $S=K3$.
The easiest way to see this is to consider a special
case of $S$ namely
$$S={\bf P}^1\times {\bf P}^1=
{T^2\over {\bf Z}_2}\times {T^2\over {\bf Z}_2}={T^4/{\bf Z}_2\over
{\bf Z}_2}={K3\over {\bf Z}_2}$$
where the last ${\bf Z}_2$ acts as an inversion on one of the
$T^2$'s.  It may appear that we are breaking supersymmetry
by taking the last $Z_2$ quotient.  This is not necessarily
the case, because this ${\bf Z}_2$ can act on the normal
direction to $S$ in
a way preserving the supersymmetry, which does not affect the geometry of $S$
itself.  The effects of the last ${\bf Z}_2$ will be at most
inducing some 5-brane charges, because it leaves a number of $T^2$'s
unchanged.  This will thus not affect the 3-brane charge computed
above in the context of $K3$.  Thus the T-duality applied
to this case should still end up giving $SU(N_f-N_c)$ as
the dual magnetic gauge group with $N_f$ hypermultiplets $ (q,\tilde q)$.
How about the magnetic Meson and the superpotential it has
with the magnetic quarks \seib ?  We do not have
a derivation of this from first principle because
we do not know the details of how the T-duality
acts on $S$.  But with a simple assumption
about how the T-duality acts, we can also recover
the Meson fields in the magnetic description:
Suppose that under the T-duality the $N_f$
3-branes that we end up with are necessarily
close to each other (this can in principle arise
if we consider the field theory limit in which
we take the volume of $S$ to be big).  If this happens
we end up getting an $SU(N_f)$ theory with 3 adjoint
hypermultiplets (the $N=4$ system on $N_f$
coinciding 3-branes) and in addition
the magnetic quarks are in the $(N_f,N_f-N_c)$ representations
of $SU(N_f)\times SU(N_f-N_c)$.  Note that the $SU(N_f)$ theory
is thus not asymptotically free, and so there is trivial infrared dynamics
associated with it.  Except that out of the three adjoint
fields of $SU(N_f)$ one of them couples to the magnetic
quarks by a superpotential term
$$W=qM{\tilde q}.$$
This follow from the $N=2$ structure of the interaction
between the 3-brane and the 7-brane (induces from the three
open string interaction representing open strings stretched
between the 3-branes ($M$) and between the 3-brane and 7-branes
$(q,\tilde q)$.
More precisely, one may consider the situation when the
gauge coupling constant of the theory on the 3-brane is small
but not zero.
The effective action for the meson field can be obtained by integrating over
rest of the fields on the 3-branes.
To the leading approximation the effective action for the meson field consists
of the
free kinetic term plus above superpotential.
In other words, the net effective theory in the infrared
is some free theory, together with an $SU(N_f-N_c)$
gauge symmetry with $N_f$ flavors in the fundamental
representation and a neutral $N_f\times N_f$ meson field
$M$ with the classical superpotential term $W=q M{\tilde q}$,
exactly as is expected in the duality proposed by Seiberg.
Note that the above mechanism of an appearance of the meson field in the
magnetic
theory is reminiscent of
how it appears in \ref\argyres{ P. C. Argyres ,  M. R. Plesser and  N. Seiberg,
``The Moduli Space of N=2 SUSY QCD and Duality in N=1 SUSY QCD'',
 Nucl. Phys. {\bf B 471} (1996) 159.}.

We now give some additional supporting evidence
to the above picture of $N=1$ duality.  Note that under
T-duality the volume of $S$ is expected to be inverted.  This
may be subject to worldsheet corrections, but the general
statement that increasing the volume of $S$ should decrease
the volume of the T-dual should be true.
Given the relation
of the volumes to the coupling constant ($V\propto {1\over g^2}$)
this is
indeed in agreement with the fact that increasing the coupling of the
electric theory should decrease the coupling of the magnetic
theory and vice-versa.

\newsec{Incorporation of Adjoint Matter}
One way to incorporate adjoint matter into the above system
is to allow $S$ to have non-vanishing $h^{1,0}$
and $h^{2,0}$. In this case we expect $h^{1,0}+h^{2,0}$
adjoint hypermultiplets \kvg .  However there will be superpotential
terms.  In particular the adjoint fields coming from $h^{2,0}$
will couple to the hypermultiplet matter, just as they
would in the case $S=K3$, where we would have an $N=2$ theory.
It is also conceivable that there would be superpotential
terms involving the adjoint $h^{2,0}$ fields, of the type
considered by Kutasov \ref\kut{D. Kutasov,
Phys. Lett. {\bf B 351} (1995) 230\semi
D. Kutasov and  A. Schwimmer, Phys. Lett.
{\bf B 354} (1995) 315\semi
 D. Kutasov ,  A. Schwimmer and  N. Seiberg,
Nucl. Phys. {\bf B 459} (1996) 455.}.  This is possible
because the zero modes of $h^{2,0}$ will have zeros along
some divisor $D$ in $S$ and this could be a potential
source for interaction.  Let us note that the dimension
of instanton moduli space in this case is given by \dimf\
$${\rm dim}{\cal M}=2N_fN_c-{\rm dim}(SU(N_c)) (1-h^{1,0}+h^{2,0})$$
We can interpret this formula in the gauge theory
setup by noting that we have an extra contribution
to the matter moduli from the adjoints coming
from $h^{1,0}$ fields which do not have any superpotential
associated with it.  On the other hand, due to the
superpotential terms associated with the $h^{2,0}$
adjoints, in the Higgs branch their expectation
value is zero. Moreover their equation of motion leads
to a constraint which should thus cut down the dimension
of Higgs branch by $h^{2,0} {\rm dim}SU(N_c)$.  This is in accord
with the above formula for the dimension of instanton moduli space.
It would be interesting to verify that not only the dimensions
match but also the moduli
spaces are identical.

We would like to thank Ken Intriligator, Sheldon Katz and Peter
Mayr
for valuable discussions.

The research of M.B., A.J. and C.V. was supported in part by
NSF grant PHY-92-18167.  The research of M.B. is in addition supported by the
NSF 1994 NYI award and DOE 1994 OJI.
The research of T.P. was supported in part by NSF grant
DMS-9500712.
The research of V.S.
was supported in part by NSF grants DMS 93-04580, PHY 92-45317
and by Harmon Duncombe Foundation.

\listrefs

\end